# Towards an Incident Management Framework in Proprietary Software Ecosystems

Luiz Alexandre Costa[a], Awdren Fontão[b], Rodrigo Pereira dos Santos[a], Alexander Serebrenik[c]

[a]*Federal University of the State of Rio de Janeiro (UNIRIO), Rio de Janeiro, Brazil*
[b]*Federal University of Mato Grosso do Sul (UFMS), Campo Grande, Brazil*
[c]*Eindhoven University of Technology (TU/e), Eindhoven, Netherlands*

## Abstract

In the evolving landscape of Software Engineering, the paradigm of software ecosystems has emerged, giving rise to proprietary software ecosystems (PSECO), with their central organizations known as keystones. PSECO is characterized by the contribution of various technologies produced as private and protected by intellectual property and confidentiality agreements, centered on common technological platforms. Sustaining these PSECO technological platforms is vital, as any incident can have substantial repercussions. This work introduces a framework for incident management to support the organizations' management teams in the PSECO context, called IM Framework. The IM Framework was developed in close collaboration with practitioners across a large international organization. We grounded the IM Framework based on the results of a rapid review study that retrieved 293 studies, of which 23 were selected after applying review procedures. This framework comprises five core categories: organizational goals, practices, success factors, associated benefits, and prevalent barriers. The IM Framework offers practical guidance for the PSECO management team, focusing on real-world applications to enhance reliability and resilience in a complex and dynamic software environment. Our study also promises to fill the gap in incident management governance by supporting the PSECO organization's management team and maintaining robust technological platforms amidst evolving business demands and market pressures.

## 1. Introduction

Software Engineering (SE) underwent a significant transformation, replacing the development of a single software product with a strategy in which several software products are integrated through a common technological platform [1, 2]. The erstwhile paradigm of an individualistic business vision has transitioned into an approach centered on interlinked networks and strategic alliances [3]. In this context, a concept of software ecosystems (SECO) emerged in the software industry. According to Manikas and Hansen (2013), SECO is the interaction of a set of actors on a common technological platform that results in a number of software solutions or services.

Proprietary software ecosystems (PSECO), a subspecies of SECO, can be characterized by the contribution of various products, technologies, and architectures (i.e., SAP) produced as private and protected by intellectual property and confidentiality agreements [5]. In the context of PSECO, the presence of a central organization, referred to as a keystone, is often observed. This organization tends to play a significant role in establishing governance policies [6], although its existence and predominance may vary depending on the specific nature and dynamics of each ecosystem. This variation includes scenarios where multiple organizations collaborate or share governance responsibilities (e.g., licensing model, partnership programs, and apps marketplaces), adapting to natural changes such as business evolution and technological obsolescence [7, 8].

Maintaining a technical sustainable platform (i.e., ability to evolve with emerging technologies, maintain compatibility with different system components, and ensure operational stability) has become a priority for large organizations, as indicated in a survey performed by the Gartner Group[1]. The PSECO technological platform that underpins business initiatives is built with a variety of custom and blended technologies with multiple integration points, creating architectural complexities [9, 10]. Sustaining these complex systems to avoid outages or downtime (also known as incidents) is a concern that can have serious repercussions on keystone's image and finances [11].

To mitigate the risks of interruptions, the PSECO technological platform requires governance mechanisms related to internal and external developers, IT service providers, and IT managers [12]. Some concerns go beyond technical solutions, encompassing business and social challenges such as increasing revenue, managing knowledge and software assets, and optimizing processes for incident management [13].

---
[1]https://www.gartner.com/en/documents/4006716

The market's demand for cutting-edge solutions intensifies the pressure on organizations, accelerating work pace and transmitting urgency to IT software development teams for delivering results within tighter deadlines, impacting the synergy between business strategy and technological platform evolution at PSECO [14]. As a result, some problems arise [15, 16], such as: lack of time for complete requirements specifications; inaccurate time estimates; late projects; overspending due to rework on software artifacts; and prioritization of deadlines to the detriment of software quality. The result is a software project delivered with low quality, resulting in incidents in the organization's productive environment [17], as an example reported in 2013, when PayPal accidentally credited a man $92 quadrillion[2].

This scenario contributes to the creation of a complex environment vulnerable to failures in the PSECO, presenting challenges to developers and managers in incident management [18, 19], such as: i) developing software applications capable of achieving success while maintaining the stability of the technological platform; ii) managing governance related to the multi-stakeholder technological platform architecture; and iii) monitoring the architecture of the technological platform to ensure the quality of software applications provided to end users.

The lack of an optimized incident management process that supports developers in achieving reliability and protection of services against failures [20] may be one of the reasons to prevent the keystone management team from implementing governance strategies to evaluate decisions on the PSECO technological platform. Our study seeks to fill this gap by exploring incident management governance.

To address this challenge, we developed a framework for incident management to support the organizations' management teams in the PSECO context, called IM Framework. The IM Framework seeks to guide the management team in its approach to incident management strategies. We grounded the framework based on the results of a rapid review (RR) study that retrieved 293 studies, of which 23 were selected after applying review procedures. Our work was performed in collaboration with professionals across a large international organization.

---

[2]https://edition.cnn.com/2013/07/17/tech/paypal-error/index.html

The outcome of our work is a framework that accomplishes the following: i) presents a set of **goals** that organizations seek to achieve by implementing an incident management process; ii) explores the best **practices** that organizations may implement to achieve their incident management goals; iii) identifies **success factors** for internal and external elements that may impact the execution of incident management practices; iv) highlights the **benefits** associated with adopting incident management practices; and v) identifies the **barriers** that organizations face when trying to implement incident management goals. Our IM Framework emphasizes the idea that it is not only theoretical or conceptual but can also be effectively applied in practice to guide concrete actions and decision-making.

This work is organized as follows: Section 2 presents the background and related work; Section 3 traces the research method; Section 4 describes the results synthesis of rapid review study; Section 5 discuss the findings and the implications for researchers and practitioners; Section 6 presents the threats to validity; and finally, Section 7 concludes the study with final remarks and future work.

## 2. Background and Related Work

*2.1. Software Ecosystems*

Organizations currently work cooperatively and competitively to support new products, satisfy customer needs, and incorporate innovations. So, increasing attention is being paid to connectivity and dependency in relationships between companies [21]. In this context, there are several actors involved (e.g., suppliers, outsourcing companies, software providers, and developers) that affect and are affected by the network value creation [6].

From this new perspective, researchers created a concept to be analyzed in the software industry called software ecosystems (SECO). Manikas and Hansen [4] emphasize the idea of an environment where different actors interact and collaborate around a common technological platform to develop software solutions and services. Figure 1 shows the big picture of SECO.

A SECO classification approached from a value creation perspective can be [5]: **i) proprietary:** where the source code and other artifacts produced are protected by confidentiality agreements, as they are the products that would yield revenues to the ecosystem, e.g., platform as a service and e-commerce ecosystems; **ii) open:** where the actors do not participate to obtain direct revenues from their activity in the ecosystem, e.g., Eclipse and Apache Foundations; and **iii) hybrid:** which supports proprietary and open source contributions, e.g., iOS SECO may use proprietary strategies as app store and the source code repository to drive policies in the technological platform, and open source strategies for community engagement as tools, submission, and publication contributions. Our work focused on the proprietary software ecosystems (PSECO) context.

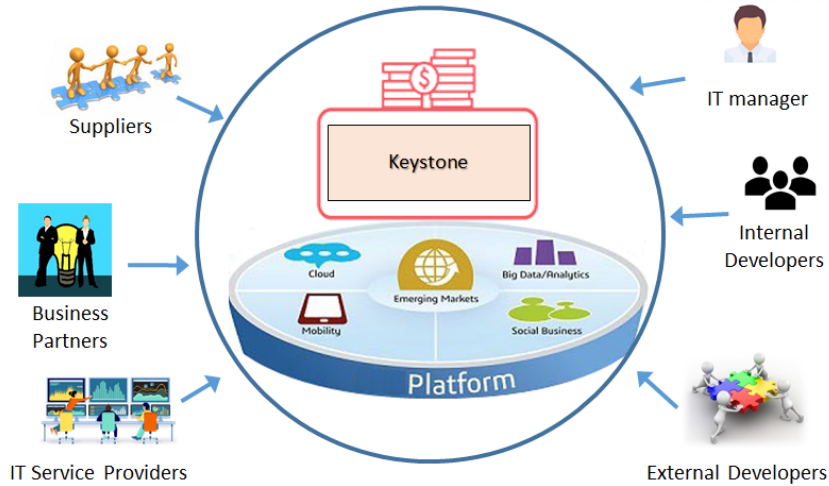

Figure 1: PSECO actors.

2.2. Incident Management in PSECO

The customized definition from ITIL[3] used in our work in the context of PSECO is: *incident management seeks not only the rapid restoration of services, but also considers the impact on software services within the ecosystem, aiming to maintain the stability of the common technology platform.*

The keystone needs to define governance strategies, such as incident management [23], to ensure the entire life-cycle of the software assets (i.e., applications, components, and services) that are part of the technology platform [24]. Incident management plays an important role in maintaining these attributes in PSECO [23]. Incidents can range from system failures to security vulnerabilities and usability issues. In this context, incident management involves the detection, triage, resolution, and communication of these incidents to ensure that end users have a positive and uninterrupted experience.

---
[3]IT Infrastructure Library (ITIL) is globally recognized as a set of best practice guidance for IT Service Management in organizations [22].

In this complex landscape of PSECO, achieving organizational objectives needs a clear alignment of strategic drivers and incident management practices [25]. Strategic drivers are the essential forces that guide an organization's path toward its desired outcomes. They encapsulate the overarching objectives, ambitions, and priorities that propel an organization forward [26]. These drivers are tailored to an organization's unique context, encompassing factors such as industry trends and internal capabilities [27].

The manner in which incident management is integrated into an organization's strategy may be customizable [28]. Different organizations, driven by their strategic goals, may identify several practices and success factors that support the management team in dealing with incident process [29]. For instance, an organization aspiring to enhance customer satisfaction may leverage incident management to swiftly address user-reported issues, thereby bolstering their user experience improvement goal. Our work emphasizes the integral role of incident management as a strategic enabler.

Different strategies related to incident management may be addressed depending on the type of SECO, such as: **i) contextual characteristics:** SECO may vary in terms of structure, governance, and business models. For example, PSECO operates under strict confidentiality policies, while open source SECO (OSSECO) favors collaboration [30]; **ii) development practices:** each type of SECO may shape its own software development practices [31]. The influence of how code is shared, updated, and maintained may affect the nature of incidents; and **iii) customized solutions:** by considering the particularities of each type of SECO, there is an opportunity to create customized incident management strategies [24].

*2.3. Related Work*

Choosing the right ecosystem strategies and governance mechanisms is a life-or-death decision for keystone and it is not a easy task [32]. The efficient development of governance mechanisms can result in a sustainable and healthy ecosystem. On the other hand, the absence of these mechanisms may lead to failure [32].

SECO governance mechanisms are managerial tools of players that have the goal of influencing an ecosystem's health [30]. The study of Alves *et* *al.* [30] pointed out that in the selection of appropriate governance mechanisms, organizations can gain strategic advantage over others leading them to better performance and to be healthier. The authors report a systematic literature review that aggregates definitions of SECO governance and classify governance mechanisms in three dimensions: value creation, coordination of players, and organizational openness and control. The results suggested that ecosystem health is under the direct influence of how governance mechanisms are implemented by organizations.

The study of Coskun-Setire *et al.* [33] investigated governance mechanisms for digital business ecosystems (DBE). The authors performed a systematic literature review that provides an overview of articles, modeling notations, modeling viewpoints, and design patterns used for DBEs. The results reinforce the need for alignment between DBE architecture and governance.

According to the study of Manikas [5], there is not much research in PSECO, also due to the difficulty of access to data from these environments. In PSECO, organizations are concerned with keeping their assets protected by intellectual property so that they are not exposed publicly, preventing this information from being used by competitors. From this challenge, the study of Costa *et al.* [31] performed a longitudinal literature study focused on PSECO governance and health to understand how the governance mechanisms works and if the organizations handle incidents on their platforms.

Finally, the study of Axelsson and Skoglund [34] investigated the challenges related to quality assurance in SECO, especially for federated embedded systems, through systematic literature mapping. The results show that the literature covered the areas of architecture, online testing, assurance practices, software operation knowledge, test suite, and review methods.

Therefore, no studies mentioned has gone deeper into the challenge of finding a way to structure the incident management strategies used in an organization's day-to-day activities to guarantee the business and process evolution in PSECO context. Our work underscores the symbiotic relationship between strategic goals and incident management in the achievement of a large international organization's objectives. As the organization defines its goals, incident management practices will be tailored to align with these objectives. By effectively managing incidents through governance and metrics, the organization not only safeguards PSECO attributes such as reliability, security, and performance but also fortifies the path to success [35]. We propose joining goals, practices, success factors, benefits, and barriers regarding the incident management in the PSECO of a large international organization to create a framework that drives the management team.

## 3. Research Method

Rapid Reviews (RR) are practice-oriented secondary studies [36] [37] [38] [39]. The main goal of a RR is to provide evidence to support practitioners' decision-making towards the solution maintaining the reliability of research protocols or removing issues that practitioners face in practice [40] - in our work, incident management in the field of Software Engineering (SE), covering PSECO. To support this goal and to meet practice time constraints, RR should deliver evidence in shorter time frames, when compared to Systematic Reviews (SR), which often take months to years [41]. In order to make RR compliant with such characteristics, some steps of SR are omitted or simplified. We followed the protocol according to the guidelines of Cartaxo *et al.* [40], similar to Kitchenham and Charters [42].

Cartaxo *et al.* delineate an RR process comprising planning, performing, and reporting phases. Our method enhances these core phases by detailing actionable steps, ensuring direct engagement of practitioners throughout each one [43]. Within our approach, Sections 3.1 and 3.2 coincide with the planning phase. Sections 3.3, 3.4, and 3.5 align with the performing phase. Section 5.9 addresses the reporting phase as proposed by Cartaxo *et al.*

### 3.1. Review Preparation

In this step, we constituted a team made up of 3 researchers and 7 practitioners who participated in the review, as described in Table 1. The RR lead researcher presented the overall objective, typical process, timeline, and expected time commitments. Next, the teams agreed on the expectations, level of involvement, and responsibilities of both researchers and practitioners.

The demand for a RR emerged from the alignment of the researcher's work-based and the industry's specific needs in a practical problem: the need to establish incident management strategies to support the IT management team, aiming at the stability of the PSECO technological platform [24].

### 3.2. Research Questions

In this step, researchers and practitioners refined the research question (RQ) as the understanding of the practitioners' context improved. To support answering the RQ, we defined five subquestions (SQ), more specifically concerning goals, practices, success factors, benefits, and barriers. This procedure was performed through an iterative process in which both teams developed an understanding of the terminology and domain concepts. In our case, it did not take much time to develop a consensus on the RQ and SQ due to the scope of each one also being grounded in academic references. Once the RQ and SQ were sufficiently clear, the research team articulated some decisions, such as the search strategy, inclusion and exclusion criteria, analysis approach (coding process).

Table 1: Characterization of RR participants.

| Qty | Profile | Skills |
| --- | --- | --- |
| 2 | Junior Systems Analyst | Initial experience (1 to 3 years) in software development. |
| 2 | Mid-Level Systems Analyst | Intermediate experience (3 to 6 years) in software development. |
| 3 | Senior Systems Analyst | Significant experience (6+ years) in software development. |
| 1 | Research Leader | RR research leader (Master's degree) with several contributions in SE. |
| 1 | Senior Research | RR research supervisor (Ph.D. degree) with notable publications in SE. |
| 1 | Ph.D. Professor | Head of SE Lab with over 15 years of experience. |

In line with our research method, our RQ aimed to allow the researcher to obtain detailed and rich information about incident management aspects of the topic at hand. The research questions evolved as we collected and examined our data, and they were further revised. Our main RQ was *"How can organizations develop an approach to incident management in their PSECO?"*.

It is worth noting that researchers and practitioners teams formulated the RQ and SQ to provide insights that can improve incident management, considering PSECO context and particularities. Each proposed SQ was structured to contribute to the building of the IM Framework. To answer all the SQ, we performed the procedures shown in Table 2. SQ were also defined in close collaboration with practitioners throughout virtual and in-person meetings moderated by the researcher.

Table 2: SQ rationales to support the rapid review study.

| SQ | Rationale |
| --- | --- |
| 1. Which are the common goals defined by organizations regarding incident management in PSECO? | Understanding of the priorities and common goals that organizations seek to achieve when managing incidents in PSECO [44, 45]. |
| 2. Which are the practices for achieving the goals defined by organizations regarding incident management in PSECO? | Providing practical guidelines adopted by organizations to achieve the goals [46, 47]. |
| 3. Which internal and external factors are identified as critical to the success of incident management in PSECO? | Identifying internal and external factors that aim to ensure the use of practices regarding incident management practices in PSECO [48, 6]. |
| 4. Which benefits are associated with the adoption of incident management practices in PSECO? | Identifying the tangible and intangible benefits that organizations can reap by adopting incident management practices in PSECO [49, 50]. |
| 5. Which barriers do organizations face when trying to implement incident management practices in PSECO?" | Providing insights into obstacles that may emerge when trying to implement incident management goals in PSECO [51, 6]. |

*3.3. Search Process*

We ground our work on Empirical Software Engineering (ESE) guidelines [52]. We performed a similar SR protocol and the procedures were described in Section 3.4. Following the RR methodological characteristics [40], we abbreviate the search for primary studies and conduct the RR under the agreed time frame. We used only the Scopus[4] search engine. It searches in many of the most relevant digital libraries. We tested many different versions of the search string until we found a set that returned relevant studies. Before conducting the search, we present the possible search string to other two

---

[4]https://www.scopus.com/home.uri

experienced researchers in Empirical Software Engineering, and through a feedback loop with them, we refined and defined the following search string:

(( "software ecosystem*" OR "software supply network" OR "software vendor*" OR "software supply industry" OR "information system*" ) AND "incident manag*" )

The extraction and analysis of data from the selected studies were carried out by two researchers. Several discussion meetings were held to clarify some doubts that required double checking the results. A third researcher validated the final set of studies.

### 3.4. Inclusion and Exclusion Criteria

We adopted the following inclusion criteria to select studies: (i) studies written in English, ii) studies must present evidence based on scientific empirical methods (e.g., interviews, surveys, case studies, etc.), and (iii) studies that answer at least one SQ. The exclusion criteria adopted in this work were: (i) secondary studies (e.g., systematic mapping studies and systematic reviews), and (ii) duplicate reports of the same study.

As first step, the literature collection started with 293 studies retrieved from the Scopus digital library. The automatic search was conducted between July and September, 2023. In the second step, we removed studies that satisfied our exclusion criteria, reaching 287 studies. In the third step, we excluded studies based on titles and abstract that did not satisfy our inclusion criteria, obtaining 255 studies. In the fourth step, we read the full text of the studies and removed those that could not answer at least one SQ, obtaining 32 studies. Finally, the research team conducted a quality assessment [53] over each study taking the practitioner's perspective and context into consideration. We selected 23 studies for data extraction and the complete list was enumerated from S1 to S23 in Appendix A. The selection and data extraction steps of the rapid review were shown in Fig. 2.

### 3.5. Coding Process

We performed an open coding approach where we coded the selected studies inspired by the initial procedures for the grounded theory of Strauss and Corbin [54]. The method uses an inductive approach (bottom-up), which means that theories emerge from collected data rather than being imposed by the researcher [55].

Two researchers conducted and coded the selected studies over an average of three iterative cycles. The codes were generated inductively, i.e., emerged directly from the data without prejudice or prior assumptions by the researchers. To ensure a collaborative, comprehensive, and reliable approach, the results were reviewed by each researcher to allow for the exchange of perspectives and consensus on code generation. During the coding process, the points of disagreement between researchers were discussed transparently and resolved through consensus in a new review meeting. One researcher with more than 15 years in ESE double-checked the results and ensured the compliance of the final dataset.

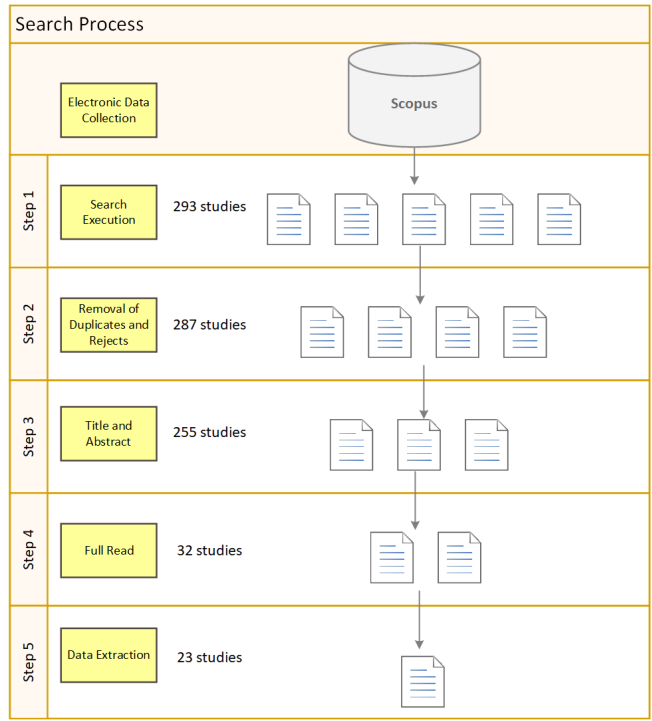

Figure 2: Selection and data extraction steps of the rapid review.

The two researchers divided the **transcripts** of the selected studies into coherent units (sentences, paragraphs, keywords, or concepts) and added **preliminary codes** that represented the key points to each SQ rationale. Based on the preliminary codes, we set the **focused codes** that captured the most frequent and relevant elements affecting incident management. Next, we formed axial coding as described by Charmaz [55] to group the codes into core categories using Atlas.TI tool. We also perform three iterative cycles with discussions among the researchers to write memos for the codes and categories and build relationships across the codes, forming a logical network to a higher degree, as shown in Table 3.

Table 3: Illustration of the coding process.

| **Transcript Unit:** ITIL covers several IT service management processes, including incident management, which is responsible for correcting failures and restoring the normal service operation, as soon as possible, minimizing the impact on business. One of the most relevant monitoring indicators related to this process is the completion time for incident resolution. **(S4)** |||
|---|---|---|
| **Preliminary Code:** Assertive and reliable estimate for completion time is still challenging. | **Focused Code:** Reduce incident resolution time | **Core Category:** Goals (SQ1) |

## 4. Results

SQ in RR are as important as in SR [42]. However, there is a subtle difference. While SR SQ are intended to identify research gaps and provide broader insights to the research community, RR SQ are more restricted, aimed at providing limited answers to the practical context in which they are embedded [40]. Therefore, once RR SQ are defined, all effort is towards answering them. In RR, results are considered useful when they help practitioners to solve or mitigate the practical problems [40].

Moreover, we propose a visual representation of SQ through a metamodel. A metamodel defines structures and meanings for the elements in a model and has become a technique to handle the complexity issues in the software development industry [56]. In our context of IM Framework, the metamodel provides how the elements of the SQ (goals, practices, factors, benefits, and barriers) are connected and interact with each other in a more structured and abstract way, as shown in Fig. 3. For example, the association among goals and practices or among benefits and practices can be visually highlighted, making it easier to understand how these elements associate each other.

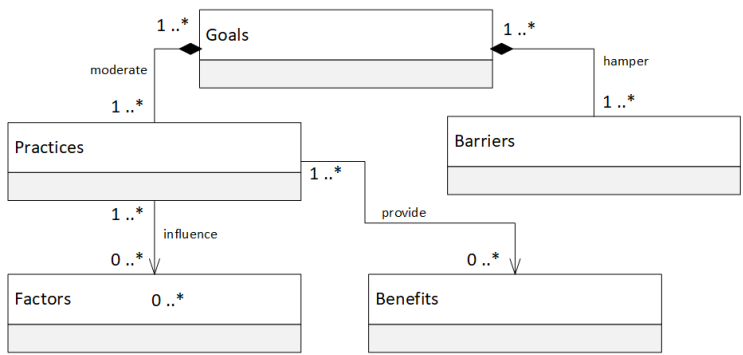

Figure 3: IM Framework metamodel with SQ elements.

*4.1. SQ1: Which are the common goals defined by organizations regarding incident management in PSECO?*

SQ1 seeks to shed light on the objectives that organizations aim to achieve when dealing with incident management in PSECO. This investigation was motivated by the need to clearly understand the goals that guide the actions of organizations to handle incidents. By exploring the literature and synthesizing the findings, it was possible to identify a comprehensive set of common goals established by organizations. These goals not only provide valuable insight into organizations' aspirations but also serve as a foundation for incident management strategies in PSECO. Table 4 presents the most frequently identified goals, revealing the shared goals that shape organizations' approach to incident management in PSECO.

Table 4: Organization's goals for incident management.

| ID | Goals | Study |
|---|---|---|
| G1 | Reduce incident response time | S1, S5, S7, S11, S12, S13, S18, S19, S22, S23 |
| | It seeks to reduce the interval between the time an incident is detected and the time a response action is initiated. | |
| G2 | Reduce incident resolution time | S1, S4, S6, S8, S12, S13, S14, S15, S16, S17, S19, S20, S21, S22 |
| | It seeks to reduce the interval between the time an incident is identified and the time it is resolved or restored. | |
| G3 | Reduce the number of incidents in the backlog | S2, S3, S10, S19, S23 |
| | It seeks to reduce the number of incidents pending resolution over time. | |

We found three goals that were somehow directly or indirectly mentioned, as follows: i) reduce incident response time: the focus is on decreasing the amount of time it takes for the sustaining team to recognize and begin to deal with an incident after it is reported; ii) reduce incident resolution time: the focus is on decreasing the amount of time it takes to fully resolve the incident; and iii) reduce the number of incidents in the backlog: the focus is on decreasing the number of incidents that are unresolved in the backlog.

Tello-Oquendo *et al.* [S1] discussed how organizations should have plans and procedures for dealing with information security incidents and how an incident management capability can help with effective incident detection, response, and recovery. This is directly associated with reducing the time needed to resolve incidents. While the paper mentions the importance of minimizing the occurrence of incidents, it focuses more on effectively managing incidents that do occur, which is in line with reducing resolution time. Fuada [S2] addressed the goal of reducing backlog incidents. Analysis of the COBIT framework for incident management in higher education institutions in Indonesia reveals a gap in the maturity of incident handling, where the level of maturity found is below expectations.

Nogueira *et al.* [S3] described a set of tools for collecting, storing, and analyzing data from software repositories, including bug-tracking information and incident management tools such as SVN, SonarQube, and Jenkins. By collecting this data, the objective is to obtain insights to improve the quality of software development processes and products, which includes the identification and resolution of incidents. The reduction in the number of incidents in the backlog is one of the expected results of this study, since the analysis of historical data can help prevent future problems and optimize the incident management process.

Amaral *et al.* [S4] discussed the challenge of estimating the ticket completion time in incident management. This is directly related to the time of response, as organizations aim to respond to incidents as quickly as possible to minimize the impact on business. The paper explores methods to improve the accuracy of these time estimates.

Raharjana *et al.* [S5] discussed some reasons linked to reduced response time: incidents take a long time to respond; adoption of best practices to manage incidents based on the COBIT framework; and evaluating the user experience. A similar approach was addressed by Palilingan *et al.* [S7]. Both studies claim that the incidents have not been a major concern for academic information systems.

Silva *et al.* [S6] presented an approach based on machine learning to automate the classification of an incident consisting on analyze descriptions written in natural language. The authors presented that incident management process requires a correct categorization to attribute incident tickets to the right resolution group aiming to have the lowest impact on the business. In summary, the study focused on the goals of reducing the time of resolution. Belov *et al.* [S8] proposed to use a management subsystem for the identification and classification of incidents through mathematical model algorithms. The authors pointed out that the effectiveness of the incident management process is determined by the speed of incident resolution.

Samopa *et al.* [S10] highlighted the importance of analyzing the root cause of incidents. Building a knowledge base to identify major incidents that frequently occur may contribute to reducing the number of incidents in the backlog. The paper [S11] addressed a problem related to the lack of visibility in IT service management processes, specifically in relation to incident management. This issue can affect the organization's response time to incidents as they cannot effectively determine how the process is working.

Goby *et al.* [S12] investigated the benefits of business intelligence methods to automate the process of ticket assignment and thus make implicit knowledge explicit. This predictive analysis can help identify ways to accelerate resolution and reduce response time. Chen and Wan [S13] discussed the importance of fast recovery from incidents to ensure users can resume their work as soon as possible. Similar to the response time goal, the paper focuses on improving incident management processes to ensure efficient coordination of IT resources and a rapid return to normal service levels. In summary, the study primarily aligns with the goals of reducing response time and resolution time.

Assuncao *et al.* [S14] discussed incident management in large IT service providers that handle customers' IT infrastructure. The authors focused on optimizing ticket dispatching policies to ensure that high-severity tickets are resolved within their service level agreements (SLA). The paper most directly relates to the goal of reducing the time to resolution.

Kundu *et al.* [S15] addressed the application of Monte Carlo simulation to determine the capability of the incident management process in an IT support organization. The study deals with the efficiency and effectiveness of the incident management process, which are linked to reduce resolution time. Bartolini *et al.* [S16] focused on presenting an approach based on scenario analysis tools to improve the performance of IT support organizations, with an emphasis on optimizing incident resolution time.

Tchoffa *et al.* [S17] discussed the causes of dysfunctions and problems that occur in distributed systems that are characterized by heterogeneity and comprise several interdependent applications whose programming is done by separate teams without communication between them. The authors pointed out the importance of incident management to restore service as soon as possible. Bartolini *et al.* [S18] focused on IT managers who need comprehensive decision support tools that enable them to analyze incident management operations, both at the level of the entire organization and at the level of the single support group. The study suggests that the tools can be used to optimize the organization's incident response capability.

Silva *et al.* [S19] reported that the main reason for ITIL implementation project failures is people's resistance to change. The authors suggest that organizations can have the best and most streamlined processes ever designed, but if the people don't have the skills to execute them, the processes are useless, and vice versa. Improving process maturity can help achieve the goals of reducing incident response time, resolution time, and backlog.

Pereira *et al.* [S20] proposed a maturity model to assess an ITIL implementation. The authors addressed the fact that the main problem resides in the fact that ITIL dictates to organizations "what they should do" but is not clear about "how they should do it". Reducing incident resolution time is a common goal related to incident management efficiency in this study.

Muhren *et al.* [S21] dealt with the investigation of how organizations that operate in high-risk environments, known as High Reliability Organizations (HRO), can influence and improve incident management in more conventional organizations. When the authors mentioned the importance of learning from incidents, they provided insights that may lead to best practices and, consequently, a reduction in resolution time.

Van Den Eede *et al.* [S22] presented a dynamic model of the performance of an organization's incident management process as determined by the capability of its supporting emergency response information system. The authors proposed concepts of adaptability, control, implicit knowledge, and explicit knowledge in order to achieve improvement in the incident management process. These concepts may affect response and resolution times.

Bandara *et al.* [S23] considered a detailed case narrative on how a leading Australian Finance organization has utilised contemporary Business Process Management (BPM) concepts for improving the IT incident management processes within the whole organization. Improving these processes can lead to faster incident responses. The implementation of good practices, such as those mentioned in the article, can indirectly contribute to the reduction of the backlog since it improves the handling of incidents.

Therefore, based on the results obtained in our SQ1, we can conclude that organizations operating in PSECO environments share different goals and priorities with regard to incident management. Among the most prominent goals, **reducing incident response time** (10 of 23 studies - 43%) and **reducing incident resolution time** (14 of 23 studies - 60%) emerge as common goals in almost two-thirds of the reviewed studies. This indicates a significant emphasis on agility and efficiency in responding to disruptions. In addition, **reducing the number of incidents in the backlog** (5 of 23 studies - 21%) is also an important concern for a considerable portion of organizations, highlighting the importance of proactively managing the backlog of unresolved issues. These results illustrate the diversity of strategic objectives that organizations seek to achieve through incident management in PSECO, highlighting the need for adaptable approaches to ensure the resilience of these environments.

*4.2. SQ2: Which are the practices for achieving the goals defined by organizations regarding incident management in PSECO?*

As we delve deeper into the landscape of incident management within PSECO, our quest to unravel best practices takes center stage. The motivation behind this inquiry lies in the practicality and relevance of understanding the practices and approaches adopted by organizations to attain their incident management objectives. The results are summarized in the Table 5.

Table 5: Practices for incident management in PSECO.

| ID | Practice | Study |
|---|---|---|
| P1 | Incident response plans | S1, S4, S5, S7, S9, S10, S13, S14, S15, S16, S19, S20, S23 |
| | Develop comprehensive incident response plans that outline the procedures to be followed when incidents occur. These plans should include clear roles and responsibilities, escalation paths, and communication strategies. | |

Table 5. Continued from previous page.

| | | |
|---|---|---|
| P2 | Real-time monitoring | S3, S4, S11, S13, S15, S16, S17, S21, S22 |

Implement robust monitoring systems to continuously track the health and performance of the technological platform. Real-time monitoring can help detect incidents early and trigger timely responses.

| | | |
|---|---|---|
| P3 | Automation | S3, S4, S5, S6, S11, S12, S15 |

Leverage automation tools and scripts to streamline incident detection and resolution processes. Automation can reduce response times and minimize human error.

| | | |
|---|---|---|
| P4 | Cross-functional teams | S18, S21, S22, S23 |

Form cross-functional incident response teams that include members from IT, development, security, and business units. Collaborative teams can address incidents more effectively.

| | | |
|---|---|---|
| P5 | Knowledge base | S6, S8, S10, S12, S14, S15, S17, S21, S22 |

Maintain a knowledge base or incident repository that documents past incidents and their resolutions. This resource can be invaluable for troubleshooting similar issues in the future.

| | | |
|---|---|---|
| P6 | Post-incident reviews | S1, S3, S4, S17, S20, S21 |

Conduct post-incident reviews to analyze the root causes of incidents and identify areas for improvement in incident management processes.

| | | |
|---|---|---|
| P7 | Change management | S12, S20, S21 |

Implement robust change management processes to ensure that updates and changes to the technological platform are thoroughly tested and do not introduce new incidents.

| | | |
|---|---|---|
| P8 | Customer feedback | S20 |

Solicit feedback from customers and end-users affected by incidents to understand their experiences and make improvements accordingly.

These best practices collectively contribute to the IM Framework that helps organizations achieve their goals of minimizing disruptions, enhancing software quality, and ensuring the stability of their PSECO. However, it is important to note that the specific practices employed may vary depending on the organization's size, industry, and unique context.

Based on the analysis of the studies, the identified practices revealed that 57% (13 of 23 studies) focused on creating **incident response plans**, followed by 39% (9 of 23 studies) that addressed **real-time monitoring**. Other relevant practices include **automation** (30%, 7 of 23 studies), **cross-functional team** building (17%, 4 of 23 studies), use of **knowledge base** (26%, 6 of 23 studies), **post-incident reviews** (9%, 2 of 23 studies), **change management** (13%, 3 of 23 studies), and **customer feedback** (4%, 1 of 23 studies). These findings provide a clear view of the most common practices and highlight the prevalence of incident response and real-time monitoring strategies in the context of incident management in PSECO.

*4.3. SQ3: Which internal and external factors are identified as critical to the success of incident management practices in PSECO?*

Our attention turns to identifying the internal and external factors that play a vital role in the success of incident management in PSECO. The motivation behind this investigation lies in the need to understand the key elements that influence the adoption of critical factors that affect the course of incident management.

Internal factors generally refer to elements within the organization's direct control, such as its culture, organizational structure, human resources, and internal processes. External factors encompass aspects of the organization's external environment, such as government regulations, market changes, and technological advances. Establishing this boundary allows for a clearer approach when considering how different elements affect incident management and how the organization can prepare to deal with factors beyond its direct control. As shown in Table 6, we proposed a correlation between success factors and best practices (Section 4.2) to demonstrate that internal and external factors are guided by practices in incident management in PSECO.

Table 6: Success factors for incident management in PSECO.

| ID | Factor | Type | Studies |
|---|---|---|---|
| F1 | Organizational culture | Internal | S4, S5, S7, S16, S18, S19 |
| | A culture that promotes open communication, continuous learning, and accountability is fundamental to creating an environment conducive to effective incident management. | | |
| F2 | Qualified teams | Internal | S1, S5, S7, S9, S10, S13, S17, S21 |
| | Well-trained, experienced, and up-to-date incident response teams are essential to effectively identifying and resolving problems. | | |
| F3 | Proactive monitoring | Internal | S1, S3, S4, S6, S8, S11, S12, S13, S15, S17, S22 |
| | The ability to detect incidents in real time through monitoring systems is crucial for quickly responding to threats. | | |
| F4 | Communication with stakeholders | External | S2, S7, S9, S14, S21, S23 |
| | Effective communication with customers, end-users, partners, and stakeholders during an incident is critical to maintaining the organization's trust and reputation. | | |
| F5 | IT providers and third-parties | External | S15 |
| | Ensuring that IT providers and third parties are aligned with the organization's incident management practices is important to avoid different performance indicators and external vulnerabilities. | | |
| F6 | End-user feedback | External | S12, S20 |
| | Feedback from end-users affected by incidents can provide valuable information for continuous improvements. | | |

Therefore, based on the results obtained in our SQ3, it is evident that certain factors were frequently highlighted as critical to the success of the practices adopted by organizations operating in PSECO environments. **Organizational culture**, when addressed in 26% of studies (6 of 23), reflects its importance in creating an environment conducive to dealing with incidents, as a culture that values transparency and learning from errors facilitates effective problem solving. **Qualified teams**, mentioned in 34% of studies (8 of 23), play a crucial role, as technical expertise and specialized skills are essential for resolving complex incidents. **Proactive monitoring**, emphasized in 47% of studies (11 of 23), stands out due to its ability to identify and anticipate problems, preventing them from turning into serious incidents. **Communication with stakeholders**, addressed in 26% of studies (6 of 23), emerges as a key piece for aligning expectations and keeping all parties informed during incident management. Although on a smaller scale, the recognition of the importance of **IT providers and third-parties** in just 4% (1 of 23), as well as **end-user feedback** in 8% of cases (2 of 23), demonstrates an emerging area of research, indicating the growing need to integrate these elements into holistic incident management in PSECO context.

## 4.4. SQ4: Which benefits are associated with the adoption of incident management practices in PSECO?

By understanding the tangible and intangible gains that organizations can achieve by implementing these practices, we can assess the positive impact that effective incident management can have on the operations and overall performance of a PSECO. As shown in Table 7, we proposed a correlation between benefits and best practices (Section 4.2) to demonstrate how benefits are guided by practices in incident management in PSECO.

Table 7: Benefits for incident management in PSECO.

| ID | Benefit | Studies |
|---|---|---|
| B1 | Improve customer satisfaction | S1, S4, S5, S7, S13, S14, S15, S23 |
| | Responding quickly to incidents can improve customer satisfaction and trust. | |
| B2 | Increase operational efficiency | S3, S4, S5, S6, S7, S9, S10, S11, S12, S13, S15, S17, S18, S20, S21, S23 |

Table 7. Continued from previous page.

| | | |
|---|---|---|
| | Effectively managed incidents can result in smoother, more efficient operations. | |
| B3 | Save costs | S2, S3, S7, S10, S12, S21 |
| | Reducing downtime and operating costs can lead to direct financial savings. | |
| B4 | Improve the organization's reputation | S1, S2, S8, S9, S14, S16, S23 |
| | Handling incidents well can protect the organization's reputation in the market. | |
| B5 | Increase resilience | S1, S2, S8, S9, S10, S13, S16, S19, S21, S22 |
| | A structured approach can overcome incidents, interruptions, or failures with the least possible impact on operations. | |

Therefore, our results obtained in SQ4 revealed a diversity of benefits associated with implementing incident management practices in PSECO. Studies have consistently demonstrated that the **improvement in customer satisfaction**, present in 34% (8 of 23), is driven by rapid problem resolution, minimizing impacts on the services provided. In parallel, the **increase in operational efficiency** mentioned in 69% (16 of 23) results from the optimization of internal processes, reducing downtime, and improving productivity. Additionally, **cost savings** (26%, 6 of 23) are a direct result of effective incident resolution, reducing expenses associated with downtime and prolonged maintenance. The **improvement in the organization's reputation** (30%, 7 of 23) is seen when dealing effectively with incidents, conveying a proactive and reliable image in the face of technical problems. Finally, the **increase resilience** (43%, 10 of 23) is achieved by strengthening the organization's ability to deal with incidents, ensuring operational continuity even in the face of adverse events. These results highlight the breadth and relevance of these benefits for keystones that adopt incident management strategies in PSECO context.

*4.5. SQ5:Which barriers do organizations face when trying to implement incident management practices in PSECO?*

The barriers organizations face in implementing incident management practices can offer insights into the obstacles organizations need to overcome. Analyzing these barriers will help build strategies for successful implementation and develop solutions to mitigate these challenges in PSECO. Table 8 shows the barriers identified in this context.

Table 8: Organization's barriers (or obstacles) for incident management.

| ID | Barriers | Study |
|---|---|---|
| O1 | Lack of clear goal setting by IT board | S7, S11, S15, S16, S19, S22, S23 |
| O2 | Limited budget | S16, S17, S19 |
| O3 | Lack of qualified professionals | S1, S8, S9, S10, S13, S18, S19, S20 |
| O4 | Technological complexity | S1, S6, S7, S8, S12, S14, S16, S22 |
| O5 | Lack of adequate supporting tools | S3, S4, S5, S6, S13, S14, S15, S16, S17, S21 |

The studies (S7, S11, S15, S16, S19, S22, S23 - 30%) pointed out that, due to the **lack of clear goals**, sustaining and software projects teams may not have clear direction to understand the organization's priorities in relation to incident management. Additionally, in this context, we also included approaches that highlight the importance of measuring and evaluating the capacity of the incident management process. The studies (S16, S17, S19 - 13%) reported the importance of optimizing the allocation of financial resources to implement goals related to incident management. **Limited budget** can lead the organization to face difficulties in acquiring the necessary tools, training staff, or investing in automation. This can limit the ability to respond effectively to incidents.

The studies (S1, S8, S9, S10, S13, S18, S19, S20 - 34%) mentioned that the **lack of qualified professionals** to direct, treat, correct, and manage incidents can be a major obstacle. This field requires technical knowledge, problem-solving skills, and experience. The absence of these human resources can hamper the ability to respond to incidents.

In complex software environments, incident management can be challenging due to the **technological complexity** of interactions among applications, according to the studies (S1, S6, S7, S8, S12, S14, S16, S22 - 34%).

The architectural complexity can make it difficult to detect and resolve incidents, making the process more time-consuming and prone to recurring errors. Finally, the studies (S3, S4, S5, S6, S13, S14, S15, S16, S17, S21 - 43%) emphasized **lack of adequate supporting tools**, such as monitoring software, automation, or incident management systems. These tools are essential for facilitating incident detection, response, and documentation.

## 5. Discussion and Implications

In our analysis, each proposed SQ was structured to contribute to the building of the IM Framework. Thus, the IM Framework consists of five core categories: 1) **goals** that organizations seek to achieve by implementing an incident management process; 2) **practices** that organizations may implement to achieve their incident management goals; 3) **success factors** for internal and external elements that may impact the execution of incident management practices; 4) **benefits** associated with adopting incident management practices; and 5) **barriers** that organizations face when trying to implement incident management goals. These five core categories are key components in the framework that emerged from our analysis, as shown in Fig. 4. The IM Framework is our main research outcome and it also helped us refine our research questions as presented in Section 3.2. This section describes all categories, the main findings, and future perspectives on using the framework.

### 5.1. IM Framework - Goals

**Reduce response time (G1).** The presence of multiple developers and a common technological platform [4], combined with the need for central control and integration of external contributions, reinforces the importance of rapid incident response. Agility in identifying and resolving problems is crucial to maintaining the stability of the ecosystem and ensuring the continuity of operations between the actors involved, as noticed in 43% of the selected studies (Section 4.1).

**Reduce resolution time (G2).** The contributions and extensions provided by internal and external developers result in a pool of potentially reusable solutions which can serve as valuable assets for future development [57]. Thus, the agility in identifying and solving problems is essential to minimize interruptions and guarantee the operation of the technological platform, as mentioned in 60% of the selected studies (Section 4.1).

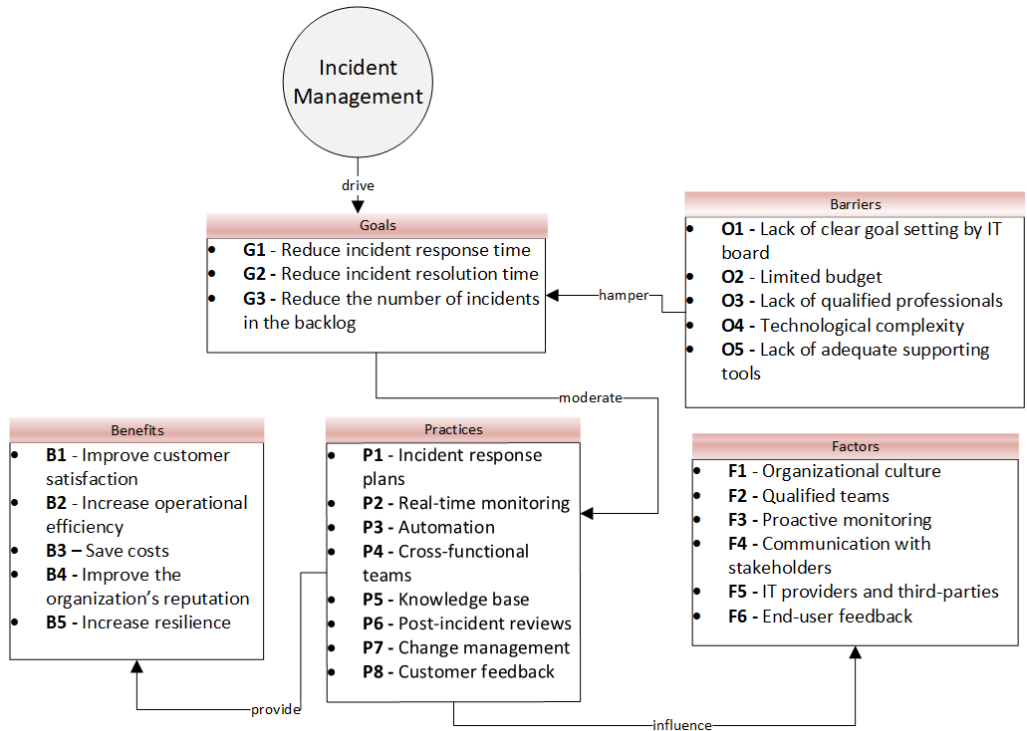

Figure 4: IM Framework overall.

**Reduce the number of incidents in the backlog (G3).** According to Pettersson *et al.* [58], PSECO relies on a common technological platform to support a specific technology and is made up of applications such as a set of reusable core components and a set of ready-to-use standard solutions . Thus, proactive actions in identifying and resolving problems contribute to avoiding the accumulation of incidents in the backlog and maintaining the stability of this technological platform, as described in 21% of the selected studies (Section 4.1).

*5.2. IM Framework - Practices*

**Incident response plans.** Structured plans ensure a unified approach to identifying, mitigating, and resolving problems, maintaining stability and confidence in the functioning of the SECO environment, as reported in 57% of the selected studies (Section 4.2). Furthermore, the centralization of these plans allows for a quick and effective response, even in a distributed and collaborative environment in which the presence of internal and external developers contributing to a shared technological platform challenges the coordination of actions between actors [57].

**Real-time monitoring.** The dynamic nature of PSECO, often featuring variability-enabled architectures [59], introduces additional challenges for monitoring. The vast number of possible configurations and combinations of components can create a vast amount of data to analyze, making it difficult to identify anomalies and isolate root causes quickly, as argued in 39% of the selected studies (Section 4.2) .

**Automation.** 30% of the selected studies (Section 4.2) declared that automating routine tasks minimizes the risk of human error, which can be detrimental to incident response efforts. Furthermore, automated processes ensure consistent and reliable execution, leading to faster resolution times and improved outcomes [60].

**Cross-functional teams.** This practice enables parallel work and faster communication, accelerating the process of identifying, analyzing, and resolving incidents. It minimizes downtime and service disruptions, ensuring the continued operation and performance of the PSECO [61], also mentioned in 17% of the selected studies (Section 4.2).

**Knowledge base.** PSECO often involves contributions from external developers and partners [62]. A shared knowledge base fosters collaboration and knowledge transfer between internal and external stakeholders. Thus, by providing a central repository of information about known issues and solutions, the knowledge base can significantly reduce the time required to diagnose and resolve incidents. This leads to faster recovery times and minimizes downtime for the PSECO, as described in 26% of the selected studies (Section 4.2).

**Post-incident reviews.** Also known as *post-mortem* reports. This practice provides a valuable opportunity for keystones to learn from their mistakes and identify areas for improvement in their incident life-cycle [63]. By analyzing past incidents and their root causes, keystones can develop and implement preventative measures to avoid similar problems in the future, as emphasized in 9% of the selected studies (Section 4.2)

**Change management.** 13% of the selected studies (Section 4.2) pointed out that by carefully planning and controlling changes, keystones can minimize the potential for disruptions and ensure the continued operation of the PSECO common technological platform. Next, by validating changes after and before they are implemented, keystones can ensure that the changes do not introduce errors or regressions into the system. This leads to higher quality and improved performance and reliability in the entire ecosystem [64].

**Customer feedback.** Integrating this practice into the IM framework for PSECO is relevant for several reasons: i) listening to and addressing customer concerns, keystones can develop products that better meet their needs and expectations; ii) demonstrating that the feedback is valued and acted upon, keystones can build stronger relationships with their customers; and iii) offering applications that are tailored to the specific needs of their customers, keystones can gain a competitive advantage. The study of Figalist *et al.* [65] also highlights the overview of the benefits of collecting and analyzing customer feedback, as well as 4% of the selected studies (Section 4.2).

*5.3. IM Framework - Factors*

**Organizational culture.** The technological platform that allows for external contributions fosters a culture of innovation and continuous improvement, according to Bosch [66]. It facilitates the coordination and collaboration among internal and external developers, leading to more efficient incident resolution, as mentioned in 26% of the selected studies (Section 4.3).

**Qualified teams.** PSECO environments often deal with complex solutions [24]. The diverse configurations and product variations within PSECO necessitate a team with the ability to analyze and troubleshoot issues across different contexts. It requires knowledge of different configurations and a flexible approach to problem-solving, as reported in 34% of the selected studies (Section 4.3).

**Proactive monitoring.** Early identification of potential issues allows for timely intervention and resolution, minimizing downtime and disruptions to the ecosystem. Furthermore, 47% of the selected studies (Section 4.3) and Santos [1] converge on stakeholder satisfaction by demonstrating that the keystone has a commitment to maintaining the common platform technological stable and reliable.

**Communication with stakeholders.** The development of software projects in PSECO context requires the collaboration of several individuals, groups, and organizations that form a network of interdependent stakeholders [67]. Thus, as also pointed out by 26% of the selected studies (Section 4.3), open and transparent communication builds trust with stakeholders and fosters collaboration during incident resolution.

**IT providers and third-parties.** 4% of studies are in line with the study of Jansen [68], in which the presence of a common platform necessitates close collaboration with IT providers and third-parties who have expertise in such technology. It ensures that these entities can support the platform and address any incident that arise.

**End-user feedback.** As described in 8% of the selected studies (Section 4.3), end-user feedback provides valuable insights into their needs, preferences, and pain points, allowing keystones to prioritize improvements and develop solutions that address real user problems. This finding converges on the study of Chung [69], who recognized user pain points to attract new ecosystem customers.

*5.4. IM Framework - Benefits*

**Improve customer satisfaction.** The common technological platform enables centralized incident resolution, ensuring consistency and reducing the risk of discrepancies across different configurations. It fosters a sense of trust and predictability for customers, leading to increased satisfaction [31]. 34% of the selected studies (Section 4.4) agreed with Costa *et al.* [31] about strengthening the position in the market.

**Increase operational efficiency.** By using shared software assets to resolve incidents across the common technological platform, keystone avoids rework, ensures consistent solutions, reduces downtime, and improves resource allocation [4, 70]. 69% of the selected studies (Section 4.4) also converge on the same understanding.

**Save costs.** By optimizing incident resolution processes and minimizing manual effort in the PSECO context (which is greater pressure for revenue [31]), keystone can free up resources for core business activities and strategic initiatives, according with 26% of the selected studies (Section 4.4). As described by Hou and Jansen [71], rework savings quantify the cost benefits derived from the reduction of system errors.

**Improve the organization's reputation.** Minimizing the impact of incidents and demonstrating a commitment to customer satisfaction reduces the likelihood of customers switching to competitors, as mentioned in 30% of the selected studies (Section 4.4). In the same vein, a positive reputation attracts new customers and fosters loyalty among existing ones, leading to increased market share and revenue growth [72, 73], which is a differential in the PSECO context.

**Increase resilience.** By proactively preventing and mitigating incidents, keystone can minimize downtime and disruptions, ensuring business continuity and avoiding costly losses in productivity. Both the 43% of the selected studies (Section 4.4) and the study of Ramezani and Camarinha-Matos [74] converged on this point.

### 5.5. IM Framework - Barriers

**Lack of clear goal setting by IT board.** Without clear incident management goals, internal and external developers, sustaining teams, and other stakeholders may work towards different objectives, hindering collaboration and progress [75]. This obstacle is also stated in 30% of the selected studies (Section 4.5).

**Limited budget.** Bosch [45] emphasizes the need for SECO to adapt to evolving technologies and changing user needs. PSECO's technological platform is particularly dynamic, requiring strategies to be capable of adapting to new configurations and potential incidents. However, financial constraints can hinder the effectiveness of incident management on the technological platform due to the basic priorities of the features and functionalities, as reported in 13% of the selected studies (Section 4.5).

**Lack of qualified professionals.** Keystone may struggle to find professionals with the necessary skillset to maintain a robust incident management process tailored to the specific needs of PSECO. Furthermore, inadequate expertise can lead to poor quality of the software, causing an imbalance in the PSECO and generating incidents [24]. 34% of the selected studies (Section 4.5) reinforce this behavior.

**Technological complexity.** PSECO rely on a common technological platform that integrates several technologies, leading to a complex and interconnected system with intricate dependencies. It can be difficult to understand, monitor, and manage, as emphasized by 34% of the selected studies (Section 4.5) and corroborated by Wnuk *et al.* [76].

**Lack of adequate supporting tools.** SECO further rely on tools, frameworks and patterns [77, 58] to manage the development and evolution of the variability-enabled architecture, the core platform, the shared core assets, and contributions. However, the lack of automation and integration capabilities can lead to increased reliance on manual work, slowing down the incident resolution process and increasing the risk of human error, as argued in 43% of the selected studies (Section 4.5).

### 5.6. Particularities for PSECO

Developing an IM Framework with particularities for PSECO is a challenging task, considering the specificities of this environment. Based on sev-eral SECO definitions [78] and SECO characteristics in the literature, the framework brings some highlights and relevance to SE and PSECO environments, which cannot be directly applied to traditional IT Service Management (ITSM). The main reason these results may not be directly applicable to ITSM is the fundamental difference between the environments. While ITSM focuses on managing IT services in organizations, incident management in PSECO deals with the complex and highly interconnected dynamics of a set of independent actors. Here are some of these particularities that fit the PSECO context inspired by the SECO general characteristics [79]:

- **Internal and external developers [62, 77]:** our framework highlights the need for practices such as multidisciplinary teams and effective communication with stakeholders. It directly relates to the presence of internal and external developers in PSECO, where collaboration between different teams is essential for rapid and effective incident response;

- **Common technological platform [45]:** by emphasizing practices such as automation and shared knowledge, our framework recognizes the importance of a common technology platform. The common technology platform is a crucial software asset in PSECO environments, where multiple solutions coexist and interact, requiring an integrated approach to incident management on a central platform;

- **Controlled central part [57]:** the emphasis on incident response plans and real-time monitoring aligns with the need for central control in PSECO environments. It ensures that, even with decentralization, incident management is coordinated and controlled, maintaining the stability of the ecosystem core;

- **Platform enabling outside contributions [10]:** recognition of practices such as post-incident reviews and customer feedback in our framework highlights the importance of incorporating external input into the core platform. This practice reflects the collaborative nature of PSECO, where external contributions can be key to improvements and incident resolution;

- **Variability-enabled architectures [80]:** our framework incorporates practices such as testing and drilling to recognize the need to

deal with variability-enabled architectures. In PSECO, where different solutions co-evolve, testing and simulations are crucial to ensure that variability does not compromise system stability;

- **Shared core assets [70]:** the emphasis on a shared knowledge base directly aligns with the presence of shared core assets in PSECO environments. This promotes collaboration and agility in incident resolution, leveraging collective knowledge to benefit the ecosystem;

- **Automated and tool-supported product derivation [81]:** the recognition of automation and support tools is in line with the need to derive products in an automated manner across PSECO. It ensures efficiency in incident management by using appropriate tools to derive solutions quickly and accurately; and

- **Distribution channel [82]:** our framework, when considering practices such as change management and customer feedback, recognizes the importance of the distribution channel. It ensures that improvements and incident resolutions are effectively distributed and communicated to end users across a diverse ecosystem.

*5.7. Main findings*

Along with the long process of building the IM Framework that began with the execution of the RR, we gathered important observations and reflections on the research topic of incident management, which we consolidated as a set of main conclusions. From them, the academic community and practitioners can identify and open new research directions aiming to contribute to ensuring better software quality deliveries, as following:

- **Increased business model change.** Organizations are experiencing rapid and frequent changes to their business models. As a result, PSECO need to keep up with the common technological platform of this evolution to meet constantly changing demands. Incidents can disrupt or undermine these changes, making it crucial to implement incident management practices that are agile and adaptable;

- **Delivery speed.** The need to deliver rapid updates, new features, and fixes in a short time is a reality for PSECO. Incidents can delay or interrupt delivery, directly affecting the common technological platform and hence the organization's ability to meet its customers' expectations. Therefore, an adaptive and rapid incident management approach is essential to minimize the impact of these events on service delivery;

- **Organizational culture.** IT infrastructure and sustaining managers who are resistant to changes from traditional to agile ITSM methods will not be able to maintain the pace of change and rapid deliveries required by business initiatives. This resistance leads to increased expenses, an increased workload, and reduced effectiveness in the organization's operations;

- **Outdated ITSM practices.** Rapid shifts in business models and the increasing pace of delivery have made bureaucratic and centralized ITSM practices outdated. IT infrastructure and sustaining managers must modernize services and embrace comprehensive approaches to service delivery to propel business initiatives;

- **Automated and integrated monitoring tools.** The capabilities of the tools required to monitor, detect, and resolve incidents are compromised by the lack of a holistic approach in complex environments, such as PSECO. The rise in intelligent automation functionality in ITSM can help organizations improve the prediction of incidents on the technological platform;

- **Integration of quality and incident management.** PSECO need a holistic approach to incident management, including the identification of root causes. It implies that software quality management must be integrated into incident management processes. Implementing robust software development practices such as rigorous testing, code reviews, static analysis, and performance monitoring can prevent many incidents before they even occur; and

- **Blameless *post-mortem* incidents.** Meeting held after the resolution of critical incidents to analyze what, how, and why they happened and take actions to prevent them in the future. This initiative focuses on learning from the incident without blaming individuals or teams.

*5.8. Implications*

In this section, we share key takeaways for researchers and practitioners as well as propose a process for leveraging the IM Framework to systematically improve incident management experience. It is worth highlighting that our framework has a continuous improvement feature. It is flexible and subject to updates as research progresses and new practices and factors emerge in incident management.

*5.8.1. Takeaways for Researchers*

As for the theoretical contribution, our work adds long-term knowledge from the incident management literature, enriches the body of knowledge, and provides an understanding of the relationships associated with incident management. Below, we list some implications for researchers:

- **Guide for empirical research.** The IM Framework provides a guide for researchers who wish to conduct empirical studies in the area of incident management in proprietary software ecosystems. They can use the identified goals, practices, factors, benefits, and barriers as starting points for their investigations, which saves time and helps define a research scope;

- **Evaluation and benchmark.** Researchers can use the IM Framework to compare results from different studies and contexts, allowing the generalization of findings across different organizations. This can contribute to the advancement of knowledge in the area;

- **Research gap identification.** By applying the IM Framework to various organizations and contexts, researchers can identify gaps in the existing literature, highlighting areas in need of additional research. This helps direct research efforts to where they are needed most; and

- **Validation and extension.** Researchers can use the IM Framework as a basis for validating or extending existing theories and models related to incident management in PSECO. The IM Framework provides a solid structure for testing hypotheses and validating concepts.

*5.8.2. Key Findings for Practitioners*

Below, we highlight some of the practical and ready-to-use implications that the IM Framework offers to practitioners seeking to optimize incident management in their organizations:

- **Process improvement.** The IM Framework can be used as a reference map to evaluate and improve your incident management processes.

Practitioners can identify missing or underutilized practices and adapt the framework to their specific needs;

- **Setting clear goals.** The IM Framework helps IT management teams set clear, measurable goals for their incident management, such as reducing response time or improving customer satisfaction. Thus, it makes it easier for teams to align with organizational goals;

- **Risk management.** By considering identified factors and barriers, professionals can anticipate and manage risks associated with incident management. They can take proactive steps to mitigate potential obstacles;

- **Practical implementation.** The practices listed in the IM Framework provide a comprehensive set of guidelines for implementation. Professionals can use them as a checklist to ensure they are covering all essential areas of incident management;

- **Performance measurement.** Professionals can use the IM Framework to establish new metrics that will facilitate measuring and monitoring the performance of the sustaining and software project teams over time;

- **Knowledge Sharing.** Professionals can use the IM Framework to share good practices with their teams and colleagues, promoting a culture of continuous learning and improvement; and

- **Integration with existing methodologies.** The IM Framework can be adapted to integrate with existing software process development and operations methodologies such as DevOps, Agile, and ITIL, making it relevant for organizations that are already following these approaches.

To instantiate the IM Framework, we propose a guide grounded in the PDCA cycle (Plan-Do-Check-Act), a widely used management model for continuous improvement [83]. The PDCA cycle can be applied iteratively over time, allowing practitioners to constantly track and adjust incident management actions, as follows: **(1) Plan:** analyze the current incident management situation in the organization and identify the main factors or barriers; **(2) Do:** implement the strategies identified in the framework and execute planned actions to improve incident management; **(3) Check:** collect relevant metrics on incident management and evaluate the effectiveness of implemented strategies to identify areas for improvement; and **(4) Act:** make adjustments to the framework as needed and implement continuous improvements to strengthen incident management.

*5.9. Result Dissemination*

The final step in the RR is to disseminate the RR results. The dissemination actions are designed to communicate the results to practitioners and researchers. According to Cartaxo *et al.*, RR must be reported in a more straight-forward way, focusing on results and recommendations [40].

We follow the guidelines of a novel approach to summarizing research findings from ESE studies called Evidence Briefing (EB) [84], as summarized in Appendix B. EB are one-paper documents that summarize the main findings of an empirical research, so researchers and practitioners can easily consume the information to support their decision making [84].

The involvement of the practitioners was crucial in critically evaluating the final EB artifact derived from our work. This direct engagement ensured that these experts refined and validated the EB artifact through their practical expertise, guaranteeing its applicability and utility within incident management in the PSECO context. Our EB artifact is available at https://doi.org/10.5281/zenodo.10428457. We also post the RR results (e.g., IM Framework, RR selected studies, and EB artifact) online on the organization's intranet website.

## 6. Threats to Validity

The reliability of the results is directly linked to the validity of the work. Every work has threats that should be addressed and considered together with the results, considering the classification proposed in [85].

Regarding **internal validity**, our concern is the researcher's bias. The results may be affected by the researcher's bias in study selection. To mitigate this threat, while the main coding was done by the two authors of this work, a third author with more than 15 years of SE research and ESE double-checked the axial coding process and the establishment of the emerging framework goals, practices, factors, benefits, and barriers.

As **constructo validity**, to reduce cost and/or time to conduct a RR, we do not use several search engines. Our study focused on the Scopus digital library. It covers a wide range of research studies due to the scope of your search engine, as evidenced in [86]. We performed a transparent process that allowed the practitioners to make their own assessments of validity.

Regarding **external validity**, our concern is the generalization to other organizations. The results obtained may be specific to incident management in the PSECO of the studied organization (Banking, Financial Services, and Insurance industry - BFSI). We addressed this threat, leaving the framework flexible if it needed adaptations to be applied to other types of organizations.

Finally, the **reliability validity** is concerned with to what extent the data and the analysis are dependent on the specific researchers [87]. In the IM Framework design step, bias in interpreting the phrases extracted from RR may affect the reliability of the results. We mitigate this threat by carrying out several cycles of reviews and discussions among the authors of this work. Conflicts on data analysis results were resolved through rounds of discussions among the authors until they reached a consensus.

## 7. Conclusion and Future Work

This work presented a RR study on incident management in collaboration with professionals across a large international organization. We constituted a team made up of 3 researchers and 7 practitioners who participated in the review. The demand for a RR emerged from the alignment of the researcher's work-based and the industry's specific needs in a practical problem: the need to establish incident management strategies to support the IT management team, aiming at the stability of the PSECO technological platform. As a result, 293 studies were retrieved, of which 23 were selected after applying the review procedures.

Our main contribution was the development of the IM framework for incident management to support the organizations' management teams in the PSECO context. Our objective was to provide an artifact that can also be applied in practice to guide concrete actions and decision-making. Our framework identifies the essential elements for successful incident management, highlighting five key categories: goals, practices, factors, benefits, and barriers. Each category plays a key role in building a solid and effective incident management strategy.

The goals served to guide what the organization wants to achieve through incident management. The practices represented the actions and procedures that the organization needs to adopt to achieve its established goals. Identifying and understanding success factors helps the organization recognize the conditions necessary for effective implementation of incident management practices. It allows the adaptation of strategies in response to internal and external factors. Benefits highlighted the rewards and advantages that the organization can obtain by successfully implementing incident management practices. It motivates the organization to invest in its incident management strategy. Barriers identified challenges and obstacles that may arise during the implementation of established goals. Knowing these barriers allows the organization to be aware of potential problems and take steps to overcome or mitigate them.

It is important to highlight that the IM framework presented in this work is not just an abstract conceptual artifact but rather a practical guide that can be instantiated and adapted by organizations of different sectors and sizes. As organizations seek to improve incident management across their PSECO, this framework provides a solid and flexible foundation to guide these initiatives. Each organization can customize its goals, practices, factors, and benefits to overcome barriers according to its specific needs. By doing so, organizations not only strengthen their incident management approach but also provide valuable support to their management team. The framework allows the management team to make more assertive decisions, optimize resources, and guarantee a safer and more resilient software environment, directly contributing to the organization's success.

As future work, we propose to investigate the new operations management philosophy called Site Reliability Engineering (SRE) that combines software development with operations. Exploring how organizations can adopt and adapt SRE principles to improve the resilience of their systems and accelerate incident recovery is a valuable direction. Another opportunity is the application of artificial intelligence and machine learning in incident management. This technological combination can revolutionize incident detection, analysis, and response. Exploring advanced algorithms to predict incidents, automate triage tasks, and even suggest response strategies based on real-time data represents an exciting field for research and development.

## Acknowledgments

The authors thank CAPES (Proc. 88887.959659/2024-00 and 88887.955223/2024-00), FAPERJ (Procs. E-26/210.688/2019 and

211.583/2019), CNPq (Proc. 316510/2023-8), and UNIRIO (PPQ 2022 & 2023) for their support.

## Appendix A. Selected Studies for the Rapid Review

**S1**. Tello-Oquendo, L., Tapia, F., Fuertes, W., Andrade, R., Erazo, N. S., Torres, J., & Cadena, A. (2019). A Structured Approach to Guide the Development of Incident Management Capability for Security and Privacy. In ICEIS (2) (pp. 328-336).

**S2**. Fuada, S. (2019). Incident management of information technology in the indonesia higher education based on COBIT framework: A review. EAI Endorsed Transactions on Energy Web, 6(21).

**S3**. Nogueira, A. F., Sergeant, E., Craske, A., Ribeiro, J. C. B., & Zenha-Rela, M. A. (2019). Collecting Data from Continuous Practices: an Infrastructure to Support Team Development. In SEKE (pp. 687-777).

**S4**. do Amaral, C. A., Fantinato, M., & Peres, S. M. (2018, September). Attribute selection with filter and wrapper: an application on incident management process. In 2018 Federated Conference on Computer Science and Information Systems (FedCSIS) (pp. 679-682). IEEE.

**S5**. Raharjana, I. K., Ibadillah, I., & Hariyanti, E. (2018, October). Incident and Service Request Management for Academic Information System based on COBIT. In 2018 5th International Conference on Electrical Engineering, Computer Science and Informatics (EECSI) (pp. 421-425). IEEE.

**S6**. Silva, S., Pereira, R., & Ribeiro, R. (2018, June). Machine learning in incident categorization automation. In 2018 13th Iberian Conference on Information Systems and Technologies (CISTI) (pp. 1-6). IEEE.

**S7**. Palilingan, V. R., & Batmetan, J. R. (2018, February). Incident management in academic information system using ITIL framework. In IOP Conference Series: Materials Science and Engineering (Vol. 306, No. 1, p. 012110). IOP Publishing.

**S8**. Belov, A. V., & Ulaeva, E. S. (2017, September). Mathematical model of incident management in the composite applications. In 2017 International Conference" Quality Management, Transport and Information Security, Information Technologies"(IT&QM&IS) (pp. 477-480). IEEE.

**S9**. Astuti, H. M., Muqtadiroh, F. A., Darmaningrat, E. W. T., & Putri, C. U. (2017). Risks assessment of information technology processes based on COBIT 5 framework: A case study of ITS service desk. Procedia Computer Science, 124, 569-576.

# Appendix B. Rapid Review Evidence Briefing

**INCIDENT MANAGEMENT FRAMEWORK IN PROPRIETARY SOFTWARE ECOSYSTEMS**

This briefing reports scientific evidence on the challenges of developing a framework for incident management to support organizations' management teams in the proprietary software ecosystem (PSECO) context.

## FINDINGS

- **Organizations are experiencing rapid and frequent changes to their business models**. Implement agile incident management practices to adapt to rapid business model changes and sustain the evolving common technological platform in PSECO.
- **Delivery speed**. An adaptive incident management approach is crucial for PSECO to mitigate disruptions and ensure timely delivery amid rapid technology updates and customer expectations.
- **Organizational culture**. Resistance to transitioning from traditional to agile ITSM methods hampers pace, increases costs, and diminishes operational efficiency, hindering adaptation to necessary business changes.
- **Outdated ITSM practices**. Bureaucratic ITSM practices hinder modern business needs. Thus, embracing comprehensive service delivery approaches is essential for adapting to rapid business model shifts and faster delivery demands.
- **Automated and integrated monitoring tools**. Intelligent automation in ITSM enhances incident prediction, addressing the complexities in environments such as PSECO by bolstering holistic monitoring, detection, and resolution capabilities within the technological platform.
- **Integration of quality and incident management**. Use of software quality practices within incident management for PSECO, mitigating incidents by adopting robust development methods such as
- **Blameless post-mortem incidents**. Post-critical incident meetings analyze incident causes, aiming to prevent future occurrences. Emphasis lies on learning rather than attributing blame to individuals or teams.

## FRAMEWORK

**Goals**
- Reduce incident response time;
- Reduce resolution time;
- Reduce the number of incidents in the backlog.

**Practices**
- Incident response plans;
- Real-time monitoring;
- Automation;
- Cross-functional Teams;
- Knowledge base;
- Post-incident reviews;
- Change management;
- Customer feedback.

**Success Factors**
- Organizational culture;
- Qualified teams;
- Proactive monitoring;
- Communication with stakeholders;
- IT providers and third-parties;
- End-user feedback.

**Benefits**
- Improve customer satisfaction;
- Increase operational efficiency;
- Save costs;
- Improve the organization's reputation;
- Increase resilience.

**Barriers**
- Lack of clear goal setting by IT board;
- Limited budget;
- Lack of qualified professionals;
- Organizational culture resistant to change management;
- Lack of adequate supporting tools.

*Who is this briefing for?*

Software engineering practitioners who want to make decisions about incident management to support organizations' management teams in the proprietary software ecosystem (PSECO) context based on scientific evidence.

*Where the findings come from?*

All findings of this briefing were extracted from the scientific studies about incident management identified on a rapid review carried out by <AUTHOR 1> et al.<<*omitted due to blind review*>>

*What is included in this briefing?*

The framework consists of five core categories: goals, practices, success factors, benefits, and barriers. These categories emerged from the analysis of scientific studies.

*What is not included in this briefing?*

Findings that are not based on scientific studies.

*To access other evidence briefings on software engineering:*

https://eseg.cin.ufpe.br/links-importantes/evidence-briefings

Figure B.5: Evidence briefing for incident management.